\documentclass[a4paper,11pt]{article}
\pdfoutput=1 

\usepackage{jheppub} 

\usepackage[T1]{fontenc} 

\title{\boldmath On physics of a highly relativistic \\ spinning particle in the gravitational field}

\author{R. Plyatsko}
\author{and M. Fenyk}
\affiliation{Pidstryhach Institute for Applied Problems in Mechanics and Mathematics,\\3-b Naukova Street, Lviv 79060, Ukraine}

\emailAdd{plyatskor@gmail.com}
\emailAdd{fenuk85@gmail.com}

\abstract{The Mathisson-Papapetrou equations are used for investigations of influence of the spin-gravity coupling on a highly relativistic spinning particle in Schwarzschild's field. It is established that interaction of the particle spin with the gravitomagnetic components of the field, estimated in the proper frame of the particle, causes the large acceleration of the spinning particle relative to geodesic free fall. As a result the accelerated charged spinning particle can generate intensive electromagnetic radiation when its velocity is highly relativistic. The significant contribution of the highly relativistic spin-gravity coupling to the energy of the spinning particle is analyzed.

\vspace{5mm}

}

\begin{document} 
\maketitle
\flushbottom

\section{Introduction}
\label{sec:intro}

Known important properties of gravitational interaction in general relativity were discovered through investigations of motions of small test bodies (particles) in gravitational field of a massive body. For example, physics of black holes was studied by consideration of world lines and trajectories of simple test particles which follow the geodesic lines in the Schwarzschild and Kerr metrics \cite{Misner,Chandra}. Here "simple" means that the particle does not possess inner structure, with inner rotation or higher multipoles.
In classical picture of the gravitational collapse of a massive object  quantum properties of the particles do not take into account.

Electrons, protons and other particles with nonzero spin which in some classical approximation can be considered as particles with inner rotation do not follow geodesic trajectories exactly. However, as it is emphasized in \cite{Misner}, in usual situations deviations of motions of the spinning test body (particles) from the corresponding geodesic motions are very small: this conclusion follows from the Mathisson-Papapetrou equations \cite{Mathis,Papa}. Just these equations, which are the generalization of the geodesic equations for description of motions of a test rotating body in general relativity, were derived for the first time in \cite{Mathis}. (This paper is absent in  refs. of \cite{Misner}, in contrast to paper \cite{Papa} which was published much later than \cite{Mathis}). However, unusual situations with spinning particles arise when their orbital velocity in the Schwarzschild or Kerr field becomes very high, close to the speed of light. Then the influence of the spin-gravity coupling on the particles orbits can be significant \cite{98,01,05,10,11,12,13,15,15a,16,17,18}. The physical reason of this situation is connected with the fact that in a frame which moves relative to Schwarzschild's or Kerr's source with the very high velocity the values of  components of the gravitational field are much greater than in  frames with low velocities. For example, as a result of strong spin-gravity action on the particle, the space regions of existence of the highly relativistic circular orbits for spinning particles in the Schwarzschild and Kerr backgrounds are much wider than for spinless particles \cite{05,10,12,13,17,18}. This fact is interesting for analysis of possible mechanism of generation of synchrotron radiation for charged spinning particles near compact astrophysical objects.

The purpose of this paper is to investigate the contribution of the spin-gravity coupling to the energy of a spinning particle moving with high velocity in Schwarzschild's field, and to obtain some estimation for electromagnetic radiation of a highly relativistic charged spinning particle. This investigations are based on the analysis of solutions of the exact Mathisson-Papapetrou equations.

The paper is organized in the following way. In sect. 2 the Mathisson-Papapetrou equations and their physical meaning are discussed. Section 3 is devoted to the analysis of the relations following from these equations in the comoving tetrads representation for Schwarzschild's metric. The dependence of the spinning particle 3-acceleration relative to geodesic free fall as measured by the comoving observer on the particle velocity in Schwarzschild's field is evaluated. For a charged spinning particle the expression for the intensity of its electromagnetic radiation caused by the acceleration is evaluated. In sect. 4 we investigate the difference in the values of energies of the spinning and spinless particles at their high velocities in Schwarzschild's field. We conclude in sect. 5.

\section{Mathisson-Papapetrou equations}

The initial form of the Mathisson-Papapetrou equations is \cite{Mathis}
\begin{equation}\label{1}
\frac D {ds} \left(mu^\lambda + u_\mu\frac {DS^{\lambda\mu}}
{ds}\right)= -\frac {1} {2} u^\pi S^{\rho\sigma}
R^{\lambda}_{~\pi\rho\sigma},
\end{equation}
\begin{equation}\label{2}
\frac {DS^{\mu\nu}} {ds} + u^\mu u_\sigma \frac {DS^{\nu\sigma}}
{ds} - u^\nu u_\sigma \frac {DS^{\mu\sigma}} {ds} = 0,
\end{equation}
\begin{equation}\label{3}
S^{\lambda\nu} u_\nu = 0
\end{equation}
where $u^\lambda\equiv dx^\lambda/ds$ is the particle's 4-velocity,
$S^{\mu\nu}$ is the antisymmetric tensor of spin, $m$ and $D/ds$ are the mass and the covariant derivative along $u^\lambda$, respectively. Here, and in the following, greek indices run through 1, 2, 3, 4 and latin indices run through 1, 2, 3; the signature of the metric (--, --, --, +) and the unites $c=G=1$ are chosen. 

Note that eqs. \eqref{1}, \eqref{2} and \eqref{3} have an important unusual feature as compare to equations in classical (nonrelativistic) mechanics which describe the propagation of the center of mass of a rotating 
body and possible changes of its angular velocity. Namely, in classical mechanics the motion of such a body is fully determined by the given initial values of the coordinates and velocity of the center of mass and the value of the angular velocity. Another situation takes place with eqs. \eqref{1}, \eqref{2} and \eqref{3}. Indeed, because the left-hand side of eq. \eqref{1} contains the term
\begin{equation}\label{4}
\frac {D^2S^{\mu\nu}} {ds^2}, 
\end{equation}
which is proportional to the second derivative of the angular velocity, the fixed initial values of the coordinates, linear velocity and only angular velocity without given initial value of the angular acceleration, in general, are insufficient for determination of a single solution of eqs. \eqref{1}, \eqref{2} and \eqref{3}. This situation cannot be changed if after using relation \eqref{3} in eq. \eqref{1} instead of \eqref{4} to write
\begin{equation}\label{5}
- S^{\lambda\mu}\frac {D^2u_\mu} {ds^2}  -  \frac {DS^{\lambda\mu}}
{ds}\frac {Du_\mu} {ds},   
\end{equation}
because expression \eqref{5} contains  the second derivative of the linear velocity and then to determine some unique solution it is not sufficient to point out only initial values of coordinates and linear velocity, without acceleration. 

More simple case, without the second derivatives in the Mathisson-Papapetrou equations, takes place when one consider the deviation of the particle motions from geodesics in the linear spin approximation.Then it is sufficient instead of eqs. \eqref{1} and \eqref{2} to deal with equations 
\begin{equation}\label{6}
m\frac {Du^\lambda}{ds} = -\frac {1} {2} u^\pi S^{\rho\sigma}
R^{\lambda}_{~\pi\rho\sigma},
\end{equation}
\begin{equation}\label{7}
\frac {DS^{\mu\nu}} {ds}  = 0
\end{equation}
(at relation \eqref{3}, it follows from \eqref{1} and \eqref{2} that $m=const$).

To avoid the terms with too high derivatives in the exact Mathisson-Papapetrou equations it was proposed to consider instead of \eqref{1}, \eqref{2} and \eqref{3} some modified equations \cite{Tulcz,Dixon}
\begin{equation}\label{8}
\frac {DP^\lambda}{ds}=-\frac {1} {2} u^\pi S^{\rho\sigma}
R^{\lambda}_{~\pi\rho\sigma},
\end{equation}
\begin{equation}\label{9}
\frac {DS^{\mu\nu}} {ds}=2 P^{[\mu}u^{\nu]},
\end{equation}
\begin{equation}\label{10}
S^{\lambda\nu} P_\nu = 0,
\end{equation}
where
\begin{equation}\label{11}
P^\nu = mu^\nu + u_\lambda\frac {DS^{\nu\lambda}}{ds}
\end{equation}
is the particle 4-momentum. An important difference in eqs. \eqref{1}, \eqref{2}, \eqref{3} and \eqref{8}, \eqref{9}, \eqref{10} consists in the form of relations \eqref{3} and \eqref{10}: because of the second term in the right-hand side of expression \eqref{11}, the vector $P^\nu$, in general, is not parallel to $u^\nu$ and relation  \eqref{3}
does not follow from \eqref{10}. (By the way, from eqs. \eqref{8}, \eqref{9} and \eqref{10} some explicit expression for the components of $u^\lambda$ through $P^\mu$ are obtained \cite{Tod}).

Often relations \eqref{3} and \eqref{10} are treated as supplementary conditions for the Mathis\-son-Papapetrou equations. 
Without any supplementary condition, these equations describe
some wide range of the  representative points \cite{Papa}
which can be in different connection with a rotating body.
To describe just the inner rotation of the body it is necessary to fix the concrete corresponding representative point. In Newtonian mechanics, the
inner angular momentum of a rotating body is defined relative to its
center of mass and just the motion of this center represents the
propagation of the body in the space.  In relativity,  the position of the center of mass of a rotating body depends on the frame \cite{M1,M2}. 
Then condition\eqref{3}, which follows from the
usual definition of the center of mass position\cite{Mash},  is common for
the so-called proper and nonproper centers of mass. (We use the
terminology when the proper frame for a spinning body is determined
as a frame where the axis of the body rotation is at rest;
correspondingly, the proper center of mass is calculated in the
proper frame.) The usual
solutions of the Mathisson-Papapetrou equations at condition \eqref{3} in
the Minkowski spacetime describe the motion of the proper center of
mass of a spinning body, whereas the helical solutions describe the
motions of the family of the nonproper centers of mass \cite{M1,M2}.
Detailed analysis of different centers of mass is presented in \cite{Costa} where it is shown that helical motions  are fully
physical in the context of the kinematical interpretation.
 
In contrast to condition \eqref{3}, relation \eqref{10} picks out a unique world line of a spinning particle in the gravitational field. However, from physical point of view eq. \eqref{10} has explicit restriction for its applications in the region of the highly relativistic motions of a spinning particle relative to the source of the gravitational field \cite{11,16a}. Different situations that arise with condition \eqref{10} for a fast spinning particle are considered in \cite{D1,D2,D3,D4,D5}.

Both at condition \eqref{3} and \eqref{10}, the Mathisson-Papapetrou equations have the constant of motion
\begin{equation}\label{12} S_0^2=\frac12
S_{\mu\nu}S^{\mu\nu},
\end{equation}
where $|S_0|$ is the absolute value of spin. 
When dealing with the Mathisson-Papapetrou equations the condition for a spinning test particle
\begin{equation}\label{13}
\frac{|S_0|}{mr}\equiv\varepsilon\ll 1
\end{equation}
must be taken into account \cite{Wald}, where $r$ is the characteristic
length scale of the background space-time (in particular, for the
Schwarzschild metric $r$ is the radial coordinate).

Equations \eqref{1} and \eqref{2} at condition \eqref{3} can be presented through the 3-component value $S_i$ \cite{15}, where by definition
\begin{equation}
\label{14}
S_i =\frac{1}{2u_4} \sqrt{-g} \varepsilon_{ikl}S^{kl},
\end{equation}
here  $\varepsilon_{ikl}$ is the spatial Levi-Civita symbol. Then eq. \eqref{2} takes the form \cite{15}
\begin{equation}
\label{15}
u_{4} \dot {S_i} + 2(\dot u_{[4} u_{i]} -
u^\pi u_\rho \Gamma^\rho_{\pi[4} u_{i]})S_k u^k
+ 2S_n \Gamma^n _{\pi [4} u_{i]} u^\pi =0,
\end{equation}
where a dot denotes differentiation with respect to the proper time $s$, and square brackets denote antisymmetrization of  indices; $\Gamma^n _{\pi 4}$ are the Christoffel symbols.

\section{Some relations following from the Mathisson-Papapetrou equations for a highly relativistic spinning particle in the Schwarzschild field}

Let us consider eqs. \eqref{1}, \eqref{2} and \eqref{3} in the linear spin approximation when according to \eqref{6} the deviation of the spinning  particle world line from the geodesic line, for which $Du^\lambda/ds=0$, is determined by the term
\begin{equation}
\label{16}
-\frac {1} {2} u^\pi \frac{S^{\rho\sigma}}{m}
R^{\lambda}_{~\pi\rho\sigma}.
\end{equation}
Because the components $S^{\rho\sigma}$ are proportional to $S_0$, according to
\eqref{13} expression \eqref{16} is proportional to the small value $\varepsilon$. It means that when the particle velocity is not very high, i.e. when the relation $|u^\pi|\gg 1$ is not satisfied, it is possible to search the solutions of the Mathisson-Papapetrou equations in the form of some small corrections to the corresponding solutions of the geodesic equations (at the condition that the values of the Riemann tensor components are not very high). Concerning the case $|u^\pi|\gg 1$ more detailed analysis is necessary. Indeed, at first glance, even when the relation $|u^\pi|\gg 1$ is satisfied and the absolute value of the expression \eqref{16} becomes much grater than at the low velocity, one can suppose that this situation is a result of the kinematic effect only, when the value of the proper time of the highly relativistic particle is much less than for a slow particle. To verify this supposition, it is appropriate to consider the value of expression \eqref{16} in the frame comoving with the particle.

For description of the comoving frame of reference we use the set of orthogonal tetrads $\lambda^\mu_{(\nu)}$, where
$\lambda^\mu_{(4)}=u^\mu$ and the relations
\begin{equation}
\label{17}
\lambda^\mu_{(\nu)}\lambda^\pi_{(\rho)}g_{\mu\pi}=\eta_{(\nu)(\rho)}, \quad
g_{\mu\nu}=\lambda_\mu^{(\pi)}\lambda_\nu^{(\rho)}\eta_{(\pi)(\rho)}
\end{equation}
takes place (here, in contrast to the indices of the global coordinates, the local indices are placed in the parenthesis; $g_{\mu\nu}$ and $\eta_{(\nu)(\rho)}$ are the metric tensor of the curved spacetime and the Minkowski tensor, respectively). Without loss in generality, we direct the first space local vector (1) along the direction of spin. Then from eq. \eqref{6} we have \cite{15}
\begin{equation}\label{18}
a_{(i)} = -\frac {S_{(1)}}{m} R_{(i)(4)(2)(3)},
\end{equation}
where  $a_{(i)}$ are the local components of the particle 3-acceleration relative to geodesic free fall as measured by the comoving observer;  $S_{(1)}$ is the single nonzero component of the particle spin. Note that the right-hand side of eq. \eqref{18} is the direct consequence of expression \eqref{16}.

Taking into account the definition of the gravitomagnetic components $B^{(i)}_{(k)}$ of the gravitational field in general relativity according to \cite{Thorne}
\begin{equation}\label{19}
B_{(k)}^{(i)}=-\frac12 R^{(i)(4)}_{}{}{}{}{}{}_{(m)(n)}
\varepsilon^{(m)(n)}_{}{}{}{}{}{}_{(k)},
\end{equation}
eq. \eqref{18} can be written in the form
\begin{equation}\label{20}
a_{(i)} = -\frac {S_{(1)}}{m}B_{(1)}^{(i)}.
\end{equation}

Let us analyze eq. \eqref{20} in the specific case when the spinning particle is moving in the gravitational field of Schwarzschild's mass. We use the standard Schwarzschild coordinates $x^1=r, \quad x^2=\theta, \quad x^3=\varphi, \quad x^4=t$ 
when the nonzero components of the metric tensor $g_{\mu\nu}$ are
\begin{equation}\label{21}
g_{11}=-\left(1-\frac{2M}{r} \right)^{-1} , \quad g_{22}=- r^2, \quad g_{33}=-r^2\sin^2\theta, \quad
g_{44}=1-\frac{2M}{r},
\end{equation}
where $M$ is the mass of Schwarzschild's source of the gravitational field.
We consider the case when the particle moves in the plane $\theta=\pi/2$ and its spin (as well as the first space local axis (1)) is orthogonal to this plane. It is convenient to orient the second space axis along the direction of the particle's motion. Then by direct calculation according to \eqref{17}, \eqref{19} and \eqref{21} we obtain
\begin{equation}\label{22}
B^{(1)}_{(2)}=B^{(2)}_{(1)}=
\frac{3M}
{r^3}\frac{u_\parallel u_\perp}{\sqrt{\gamma^2 -1}}\left(1-\frac{2M}{r}\right)^{-1/2},
\end{equation}
\begin{equation}\label{23}
B^{(1)}_{(3)}=B^{(3)}_{(1)}=
\frac{3M}
{r^3}\frac{u_\perp^2 \gamma}{\sqrt{\gamma^2 -1}},
\end{equation}
where $\gamma$ is the relativistic Lorentz factor of the moving particle as estimated by an observer which is at rest relative to the source of the gravitational field.
Let us compare the values from (\ref{22}) and (\ref{23}) at low and high velocities. When the velocity is low with $u_\parallel = \delta_1$, $u_\perp = \delta_2$, $ |\delta_1|\ll 1$, $ |\delta_2|\ll 1$, and 
$\gamma^2-1 = \Delta^2\ll 1$, where by (\ref{9})
\begin{equation}\label{24}
\Delta^2=\left(1-\frac{2M}{r} \right)^{-1}\delta_1^2 + \delta_2^2,
\end{equation}
it follows from (\ref{22}) and (\ref{23}) that
\begin{equation}\label{25}
B^{(1)}_{(2)}=B^{(2)}_{(1)}\approx
\frac{3M}
{r^3}\frac{\delta_1 \delta_2}{\Delta}\left(1-\frac{2M}{r} \right)^{-1/2},
\end{equation}
\begin{equation}\label{26}
B^{(1)}_{(3)}=B^{(3)}_{(1)}\approx
\frac{3M}
{r^3}\frac{\delta_2^2}{\Delta}.
\end{equation}
That is, at low velocity the common term $3M/r^3$ in the expressions for the gravitomagnetic components  (\ref{25})  and (\ref{26}) is multiplied by corresponding small factors:
$$
\left|\frac{\delta_1 \delta_2}{\Delta}\right|\ll 1, \quad 
\left|\frac{ \delta_2^2}{\Delta}\right|\ll 1.
$$
In the highly relativistic region, when 
$\gamma^2\gg 1$ and both $u_\parallel^2$ and $u_\perp^2$ have order
$\gamma^2$, it follows from  (\ref{22}) and (\ref{23}) that
\begin{equation}\label{27}
B^{(1)}_{(2)}=B^{(2)}_{(1)}\sim
\frac{3M}
{r^3}\left(1-\frac{2M}{r}\right)^{-1/2}\gamma,
\end{equation}
\begin{equation}\label{28}
B^{(1)}_{(3)}=B^{(3)}_{(1)}\sim
\frac{3M}
{r^3}\gamma^2.
\end{equation}
When only  $u_\perp^2\gg 1$, with $u_\parallel^2\ll u_\perp^2$, the values from (\ref{27}) are proportional to $u_\parallel$, and the values from (\ref{28}) are proportional to $\gamma^2$. In the case, when
$u_\parallel^2\gg 1$ and $u_\perp^2\ll u_\parallel^2$,  the values from (\ref{27}) and (\ref{28}) are proportional to $u_\perp$ and $u_\perp^2$, respectively. So, according to (\ref{20}), (\ref{27}), (\ref{28}) the absolute values of $a_{(i)}$ become much grater at the highly relativistic velocities of the spinning particle. It means that the smallness of $\varepsilon$ from (\ref{13}) does not lead to the conclusion about the small influence of the particle spin on its acceleration as estimated by the comoving observer. (Note that in the above considered partial case of the particle motion in Schwarzschild's field the relation $|S_{(1)}|=|S_0|$ takes place.) 

It follows from (\ref{20}), (\ref{22}), (\ref{23}) that the absolute value of the spinning particle acceleration
$$
|\vec a| = \sqrt{a_{(1)}^2 + a_{(2)}^2 + a_{(3)}^2}
$$
is determined by the expression
\begin{equation}\label{28a}
|\vec a|= \frac{3M}{r^2}
\frac{|S_0|}{mr} |u_\perp|\sqrt{1+u_\perp^2},
\end{equation}
and the vector $\vec a$ is oriented along the radial direction. According to (\ref{28a}) $|\vec a|$ does not depend on the radial component of the particle velocity and essentially depends  on its tangential velocity.
In the case of the highly relativistic motion with $u_\perp^2\gg 1$ by (\ref{28a}) we have
\begin{equation}\label{28b}
|\vec a|= \frac{3M}{r^2}\varepsilon \gamma^2,
\end{equation}
where $\gamma$ is the Lorentz factor calculated by the tangential velocity
$u_\perp$, and $\varepsilon$ is determined in (\ref{13}). 

We use expression (\ref{28b}) to estimate the electromagnetic radiation of a spinning particle which posses the electric charge $q$. Indeed, according to the known result of the classical electrodynamic the intensity $I$ of the electromagnetic radiation in the frame where the velocity of the charge particle is equal to 0 with nonzero  acceleration $w$ is given by the expression \cite{Land}
\begin{equation}\label{28c}
I=\frac{2q^2 w^2}{3 c^3},
\end{equation}
where $c$ is the speed of light. 
Inserting into (\ref{28c}) expression (\ref{28b}) as $w$ in units where $c=1$ we get
\begin{equation}\label{28d}
I=6q^2 \frac{M^2}{r^4}\varepsilon^2\gamma^4.
\end{equation}
Equation (\ref{28d}) shows that due the term $\gamma^4$ the value $I$ can be significant for some high tangential velocities even for small values of 
$\varepsilon$ and far from Schwarzschild's horizon ($r\gg 2M$).

The results presented  in this section describe the properties of the spin-gravity coupling in the proper frame of the spinning particle.
In this context the question arises: can the highly relativistic spin-gravity coupling significantly deviate trajectories of the spinning particle from the geodesic trajectories by their description in the terms of the global Schwarzschild coordinates? Different cases of the essentially nongeodesic orbits of the highly relativistic spinning particle in Schwarzschild's field are investigated in \cite{05,11,12}.

\section{Energy of a highly relativistic spinning particle \\ in Schwarzschild's field} 

According to the geodesic equations there is expression for the energy of a spinless particle with mass $m$ in Schwarzschild's field:
\begin{equation}\label{29}
E=mu_4=m\left(1-
\frac{2M}
{r}\right)^{1/2}\gamma,
\end{equation}
where $\gamma=\sqrt{u_4 u^4}$ is the relativistic Lorentz factor calculated by the particle velocity relative to the source of the Schwarzschild field, $r$ is the standard radial coordinate. That is this energy is proportional to the $\gamma$ factor, similar as in the case of the free particle motion in special relativity. Other situations arise in the case of the spinning particle motions in  Schwarzschild's field. Then by the Mathisson-Papapetrou equations the expression for a spinning particle can be written as \cite{Tod}
\begin{equation}\label{30}
E=mu_4 + g_{44}u_\lambda\frac{DS^{4\lambda}}{ds} + \frac12 g_{4\mu,\nu}S^{\nu\mu}.
\end{equation}
In contrast to (\ref{29}) the value of energy  (\ref{30}) depends not only on the initial velocity of a spinning particle and on $r$, but on the spin-gravity coupling as well. In the partial case of the radial motion of the spinning particle in Schwarzschild's field the value of its energy does not depend on the absolute value and orientation of the spin and coincides exactly with the value of energy of the spinless particle, as well as in this case the world line of the spinning particle coincides with the corresponding geodesic world line
(it is easy to obtain this result after writing eqs. (\ref{1})--(\ref{3}) at condition $\theta=const$, $\varphi=const$). However, any nonzero value of the particle tangential velocity leads to some deviation of the value of the spinning particle energy from the value of the energy of the spinless particle. Naturally, when $|u_\perp|\ll 1$, i.e. for low values of the tangential velocity, this deviation is small. It is interesting to investigate the dependence of the spinning particle energy on the tangential velocity in the highly relativistic region when $|u_\perp|\gg 1$. For this purpose it is convenient to deal with the exact Mathisson-Papapetrou equations in the form of the first order differential equations for the 11 dimensionless quantities $y_i$ where by definition
$$
\quad y_1=\frac{r}{M},\quad y_2=\theta,\quad y_3=\varphi, \quad
y_4=\frac{t}{M},
$$
$$
y_5=u^1,\quad y_6=Mu^2,\quad y_7=Mu^3,\quad y_8=u^4,
$$
\begin{equation}\label{31}
    y_9=\frac{S_1}{mM},\quad y_{10}=\frac{S_2}{mM^2},\quad
    y_{11}=\frac{S_3}{mM^2}.
\end{equation}
These equations are presented in the explicit form in \cite{15} as 
$$
\dot y_1=y_5, \quad \dot y_2=y_6, \quad \dot y_3=y_7, \quad \dot
y_4=y_8,
$$
$$
\dot y_5 = A_1,\quad \dot y_6 = A_2,\quad
\dot y_7 = A_3,\quad \dot y_8 = A_4,
$$
\begin{equation}\label{32}
 \dot y_9 = A_5,\quad \dot y_{10} = A_6,\quad
\dot y_{11} = A_7,
\end{equation}
where $A_i$ are the corresponding functions of $y_i$ and contain the two constants of motion: the energy and angular momentum
(a dot denotes the usual derivative with respect to the dimensionless value $x=s/M$). At the fixed initial values of $y_i$ different values of these constants correspond to motions of different centers of mass of the spinning particle.

Let us consider eqs. \eqref{32} in the partial case of the spinning particle motion in the plane $\theta=\pi/2$ with the spin orthogonal to this plane. It means that in notation \eqref{31} we put $y_2=\pi/2,\quad y_6=0,\quad y_9=0,\quad y_{11}=0$ and others nonzero functions $y_i(x)$ can be find by the numerical integration of eqs. \eqref{32}. The important point in this procedure is finding values of the energy and angular momentum which correspond just to the proper center of mass of the particle for the fixed initial values of $y_i$, 
not to the helical solutions. For this purpose we use computer searching. 
As a typical case here we present the results for the spinning particle with
$$
\varepsilon_0\equiv \frac{S_0}{mM}= 10^{-2}
$$
(note that in contract to $\varepsilon$ from \eqref{13}, the value of $\varepsilon_0$ does not depend on  $r$)
which begins motion from the position $r=2.5 M$ with the initial value of the radial velocity $u_\parallel = - 10^{-2}$ with different initial values of the tangential velocity $u_\perp$. Table 1 describe the situations when the sign of the $u_\perp$ is positive with the orientation of the particle spin when $S_2\equiv S_{\theta} >0$. Table 2 corresponds to the cases with
$u_\perp (0)<0$ and the same value of the $S_2$ as in table 1. Both tables 1 and 2 show the ratio of the energy of the spinning particle $E_{spin}$ to the value of the energy of the spinless particle $E_{geod}$ which moves along the geodesic lines and start with the same initial velocity as the spinning particle. For $u_\perp (0)=0$ we have $E_{spin}/E_{geod}=1$,
exactly. When $|u_\perp (0)|\ll 1$ the value $E_{spin}$ is almost equal to
$E_{geod}$ with high accuracy. Other situations arise when $|u_\perp|$ is growing up to the highly relativistic motions with $u^2_\perp\gg 1$. According to tables 1 and 2 at the highly relativistic regime the difference between $E_{spin}$ and $E_{geod}$ is growing significantly with growing
$|u_\perp|$. There are essential difference of the data in tables 1 and 2: in the first case $E_{spin}/E_{geod}>1$ whereas in the second case $E_{spin}/E_{geod}>1$. This property corresponds to the known result that at $S_2>0$ and $u_\perp (0)>0$ the spin-gravity coupling acts on the particle as some attractive force whereas at $S_2>0$ and $u_\perp (0)>0$
this action is repulsive \cite{12,15}.

So, the contribution of the spin-gravity coupling to the energy of a spinning particle in Scwarzschild's field becomes large  when its
velocity is highly relativistic.

\begin{table}[tbp]
\centering
\begin{tabular}{|c|c|}
\hline
$u_\perp (0)$ & $E_{spin}/E_{geod}$ 
\\
\hline 
0 & 1\\
5.88 & 1.06  \\
11.75 & 1.24 \\
17.67 & 1.63 \\
23.50 & 2.00 \\
\hline
\end{tabular}
\caption{\label{tab:i} On comparison of the energies of the spinning and spinless particles at different orbital velocity for $u_\perp (0)>0$.}
\end{table}

\begin{table}[tbp]
\centering
\begin{tabular}{|c|c|}
\hline
$u_\perp (0)$ & $E_{spin}/E_{geod}$ 
\\
\hline 
-2.35 & 0.99\\
-4.70 & 0.96  \\
-11.75 & 0.76 \\
-17.62 & 0.44 \\
-21.15 & 0.20 \\
\hline
\end{tabular}
\caption{\label{tab:i} On comparison of the energies of the spinning and spinless particles at different orbital velocity for $u_\perp (0)<0$.}
\end{table}


\section{Conclusions}

In addition to the known results concerning the influence of the spin-gravity coupling on world lines and trajectories of the highly relativistic spinning particle in the Scwarzschild field \cite{12,15}, in this paper we present the results about the effect of the highly relativistic spin-gravity coupling on the particle's energy. Depending on the correlation of the spin orientation and the particle orbital velocity the values of the spinning particle energy can be much larger or less than the corresponding values for the spinless particle.

In the case of highly relativistic motions of the charged spinning particle in Scwarz\-schild's field, in sec. 3 we considered the intensity of the energy of its electromagnetic radiation as estimated in the proper frame of the particle.
It is important that this value is proportional to the $\gamma^4$, i.e. becomes very large for highly relativistic orbital velocities of the particle.

In further investigations it is interesting to apply the results of this paper to the analysis of possible role of the highly relativistic spin-gravity coupling in the astrophysical processes with fast spinning particles in strong gravitational fields.



\acknowledgments

This work was supported by the budget program of Ukraine "Support for the development of priority research areas" (CPCEC 6451230).


\end{document}